\documentclass[aps,prd,showpacs,floatfix,nofootinbib,twocolumn,10pt]{revtex4}

\usepackage{color,amsmath,amssymb,graphicx,latexsym,subfigure}
\usepackage{threeparttable,multirow,txfonts}

\begin{document}

\title{Stringent constraints on the light boson model with supermassive black hole spin measurements}

\author{Lei Zu$^{1,2}$, Lei Feng$^{1,3}$\footnote{fenglei@pmo.ac.cn},
Qiang Yuan$^{1,2,4}$\footnote{yuanq@pmo.ac.cn},
Yi-Zhong Fan$^{1,2}$\footnote{yzfan@pmo.ac.cn}}

\affiliation{$^1$Key Laboratory of Dark Matter and Space Astronomy, Purple
Mountain Observatory, Chinese Academy of Sciences, Nanjing 210033, China\\
$^2$School of Astronomy and Space Science, University of Science and
Technology of China, Hefei, Anhui 230026, China\\
Joint Center for Particle, Nuclear Physics and Cosmology,  Nanjing University -- Purple Mountain Observatory,  Nanjing  210093, China\\
$^4$Center for High Energy Physics, Peking University, Beijing 100871, China
}

\begin{abstract}
 Massive bosons, such as light scalars and vector bosons,
can lead to instabilities of rotating black holes by the superradiance
effect, which extracts energy and angular momentum from rapidly-rotating
black holes effectively. This process results in spinning-down of
black holes and the formation of boson clouds around them.
In this work, we used the masses and spins of supermassive black holes
measured from the ultraviolet/optical or X-ray observations to constrain
the model parameters of the light bosons. We find that the mass range of
light bosons from $10^{-22}$ eV to $10^{-17}$ eV can be largely excluded
by a set of supermassive black holes (including also the extremely massive
ones OJ 287, Ton 618 and SDSS J140821.67+025733.2), particularly for the
vector boson scenario, which eliminates a good fraction of the so-called
fuzzy dark matter parameter regions.
For the scalar bosons with self-interaction, most part of the mass range
from $\sim 3 \times 10^{-19}$ eV to $10^{-17}$ eV with a decay constant
$f_a > 10^{15}$ GeV can be excluded, which convincingly eliminate the QCD
axions at these masses.
\end{abstract}


\maketitle

\section{Introduction}

The scattering between a rotating black hole (BH) and light bosons can
extract energy and angular momentum of the BH, which is the so-called
superradiance effect \cite{1969NCimR...1..252P,1972ApJ...178..347B,
1972PhRvL..28..994M,1972Natur.238..211P,1973ApJ...185..649P,
1980PhRvD..22.2323D,2012PhRvL.109m1102P,2015PhRvD..91h4011A,
2015LNP...906.....B,2011PhRvD..83d4026A,2017PhRvD..96b4004E,
2018PhRvD..98h3006S}. This phenomenon takes place when the superradiance
condition is satisfied, i.e.,
\begin{equation}
0<\omega<mw_{+},
\label{eq:1}
\end{equation}
where $\omega$ is the frequency corresponding to the light boson, $m$ is
the magnetic quantum number, and $w_{+}$ is the angular velocity of the
BH horizon which reads
\begin{eqnarray}
w_{+}=\frac{1}{r_{g}}\frac{a_{*}}{1+\sqrt{1-a_{*}^2}},
\label{eq:2}
\end{eqnarray}
where $a_{*}=a/r_{g}$ is the dimensionless spin parameter of the BH with
$a=J/M_{\rm bh}$ being the spin-to-mass ratio and $r_{g}=GM_{\rm bh}$ being
the gravitational radius.

When the wavelength of the light boson is comparable to the size of the BH,
the number of bosons surrounding the BH grows exponentially to form a boson
cloud. The self-interaction of the bosons would lead to the collapse of
the cloud when reaching a critical size, which is known as ``bosenova''
\cite{2001Natur.412..295D,2011PhRvD..83d4026A}. The superradiance process
extracts energy and angular momentum from the BH, which enables us to
constrain the parameters of the light boson if the masses and spins of
BHs have been reasonably measured \cite{2011PhRvD..83d4026A,
2019PhRvL.123b1102D,2019arXiv190802312N,2019arXiv191107862F}.

Supermassive black holes (SMBH) at the centers of galaxies have been found
with masses $\sim 10^6-10^{11}~M_{\odot}$. To measure the spin of the SMBHs
is somehow challenging. By means of the first direct imaging of the SMBH in
the center of M 87 with the Event Horizon Telescope (EHT), the mass of the
SMBH was determined to be $\sim 6.5 \times 10^9$ M$_{\odot}$
\cite{2019ApJ...875L...1E,2019ApJ...875L...6E}, and it was inferred to be
likely highly spining \cite{2019ApJ...875L...5E,2020MNRAS.492L..22T}.
The EHT measurement was then applied to exclude the ultralight bosons with
mass between $8.5 \times 10^{-22}$ eV and $4.6 \times 10^{-21}$ eV for a given
BH time scale $\tau_{\rm bh} \sim 10^9$ years \cite{2019PhRvL.123b1102D}.
For most of SMBHs, there are lack of direct imaging of the shadows, and the
spin parameters are instead estimated through fitting the X-ray and/or
UV-optical spectral energy distributions \cite{2013AcPol..53..652B,
2015MNRAS.446.3427C,2016MNRAS.463.4041B}.

In this work we employ the currently available sample of high-spin SMBHs
based on the UV-optical/X-ray spectroscopy method \cite{2015MNRAS.446.3427C,
2019ApJ...873..101Z}, to constrain the ultralight boson model. The sample
of SMBHs with different masses can cover a wide mass range of the light
bosons. Moreover, with a few extremely massive BHs, we probe the fuzzy
dark matter (FDM) scenario with boson mass of $(1\sim10) \times 10^{-22}$~eV
\cite{2000PhRvL..85.1158H,2017PhRvL.118n1801D,2017PhRvD..95d3541H}
which is almost out of reach via only the EHT observations of
M 87 \cite{2019PhRvL.123b1102D}.

\section{Superradiance}

Superradiance is a kind of Penrose process related to waves.
A massive boson field in the Kerr background may cause an unstable
solution with an imaginary part of the frequency \cite{2011PhRvD..83d4026A,
1979AnPhy.118..139Z,2007PhRvD..76h4001D}.
It leads to an exponential growth of the number of bosons, forming a
``gravitational atom'' with energy levels ($c=\hbar=1$)
\begin{equation}
\epsilon \simeq \mu\left(1-\frac{\alpha^2}{2 \bar{n}^2}\right),
\label{eq:3}
\end{equation}
where $\mu$ is the mass of the boson, $\alpha=r_{g}\mu$, and
$\bar{n}=n+l+1$ with $n$ and $l$ being the principal and orbital quantum
numbers, respectively \cite{2011PhRvD..83d4026A,2007PhRvD..76h4001D}.



The leading contribution of the imaginary part of frequency is
(i.e., the fastest-superradiating mode of the small-$\alpha$ analytical
approximation) \cite{1980PhRvD..22.2323D,2017PhRvD..96c5019B}
\begin{equation}
\Gamma_{s}=\frac{1}{24}a_{*}r_{g}^8\mu^9,
\label{eq:4}
\end{equation}
\begin{equation}
\Gamma_{v}=4a_{*}r_{g}^6\mu^7,
\label{eq:5}
\end{equation}
where $\Gamma_s$ and $\Gamma_v$ are the superradaince rates related to
scalar and vector bosons.
In this work, we use the leading term of the analytical solution, which
is consistent with the numerical results \cite{2007PhRvD..76h4001D,
2017PhRvD..96c5019B}, to constrain the masses of the bosons.

Without taking into account self-interactions, the number evolution is simply
\begin{equation}
\frac{dN}{dt}=\Gamma N.
\label{eq:6}
\end{equation}

For a BH with given spin, the maximally allowed size of the boson cloud is
\begin{eqnarray}
N_{\rm{max}} \simeq \frac{GM_{\rm bh}^2}{\mu} \Delta a_{*},
\label{eq:7}
\end{eqnarray}
where $\Delta a_{*}$ is the difference between the initial and final spins
of the BH. The corresponding boson parameter space is simply ruled out if the
superradiance process is efficient enough that the BH lose too much
angular momentum within its lifetime $\tau_{\rm bh}$, i.e.,
\begin{equation}
\Gamma \tau_{\rm{bh}}>\ln N_{\rm{max}},
\label{eq:8}
\end{equation}
which in turn sets a bound on the boson parameters.

As the size of the cloud keeps growing,  the self-interaction effect becomes important and some nonlinear effects would be produced on this system \cite{Fukuda:2019ewf}. When it grows up to a critical size, $N_{\rm bosenova}$,
the cloud collapses which is known as ``bosenova''. For scalar particles
we have
\begin{eqnarray}
N_{\rm bosenova} \simeq 10^{78} c_{0} \frac{n^4}{\alpha^3} \left(\frac
{M_{\rm bh}}{M_{\odot}}\right)^2 \left(\frac{f_{a}}{M_{\rm{pl}}}\right)^2,
\label{eq:9}
\end{eqnarray}
where $f_a$ is the decay constant for the scalar bosons, $M_{\rm{pl}}=
2\times10^{18}$ GeV is the Planck energy, and $c_0\sim 5$ is determined
by numerical simulations \cite{2015PhRvD..91h4011A,2012PThPh.128..153Y}.

If the self-interaction is strong enough (i.e., $N_{\rm{bosenova}}<
N_{\rm{max}}$), the scalar cloud collapses when its size reaches
$N_{\rm{bosenova}}$. Thus this process would repeat
$N_{\rm{max}}/N_{\rm{bosenova}}$ times at most and the maximally allowed
size of the boson cloud would be replaced by $N_{\rm{bosenova}}$. Therefore
the exclusion condition Eq.~(\ref{eq:8}) can be revised as
\cite{2015PhRvD..91h4011A}
\begin{eqnarray}
\Gamma \tau_{\rm{bh}}(N_{\rm{bosenova}}/N_{\rm{max}})> \ln N_{\rm{bosenova}}.
\label{eq:10}
\end{eqnarray}

\section{The SMBH sample and constraints}

A widely adopted way to estimate the spin of an SMBH is to fit the radiation
spectrum in UV-optical or X-ray bands (i.e., to model the AGN continuum
emission or relativistic X-ray reflection). We summarize some SMBHs with
inferred high spins with this method in Table~\ref{table:BH}
\cite{2015MNRAS.446.3427C,2019ApJ...873..101Z}. There are some even more
massive black holes. OJ 287 has the dynamically measured mass $M_{\rm bh}
=(1.8348 \pm 0.0008)\times10^{10}$~M$_{\odot}$ and spin $a_{*}=0.381 \pm
0.004$ \cite{2018ApJ...866...11D}.
The masses of Ton 618 and SDSS J140821.67+025733.2 are estimated to be
$6.6 \times 10^{10}$~M$_{\odot}$ \cite{2004ApJ...614..547S} and $1.96 \times
10^{11}$~M$_{\odot}$~\cite{2017ApJS..228....9K}, respectively. The allowed
parameter space of the BH mass and spin within the thin disc accretion model
has been examined in Ref.~\cite{2016MNRAS.456L.109K}, giving lower limits
of the spin of about 0.6 for Ton 618 and 0.97 for SDSS J140821.67+025733.2.
These SMBHs will also be included in this study. Note that for the spins
with only lower limits available, we adopt the $90\%$ lower limits in the
calculation. Following \cite{2017PhRvD..96c5019B}, we assume approximately Gaussian errors to estimate
the $90\%$ interval on the masses. For the timescales that the SMBH maintains such a high spin,
we take the Salpeter time $\tau_{\rm{Salpeter}} \sim 4.5 \times 10^7$ years
\cite{2009ApJ...690...20S}, which is the Eddington limit for the accreting material.


\begin{table*}[!htb]
\centering
\caption {SMBHs with masses and spins measured with various method.}
\begin{tabular}{lccccc}
\hline \hline
 & $\rm{Object}$ & $M_{\rm bh}~(10^8{\rm M}_{\odot})$ & $\rm{Spin}$ & Refs. \\
\hline
&Mrk 110     & $0.251_{-0.061}^{+0.061}$ & $ 0.96_{-0.07}^{+0.03}$ & \cite{2004ApJ...613..682P,2013MNRAS.428.2901W} \\
&Mrk 335     & $0.142_{-0.037}^{+0.037}$ & $>0.91$ & \cite{2004ApJ...613..682P,2015MNRAS.446..633G} \\
&NGC 3783    & $0.298_{-0.054}^{+0.054}$ & $>0.98$ & \cite{2004ApJ...613..682P,2011ApJ...736..103B} \\
&NGC 4051    & $0.019_{-0.008}^{+0.008}$ & $ >0.99$ & \cite{2004ApJ...613..682P,2012MNRAS.426.2522P} \\
&NGC 4151    & $0.457_{-0.047}^{+0.057}$ & $>0.90$ & \cite{2006ApJ...651..775B,2015ApJ...806..149K} \\
&NGC 5506    & $0.051_{-0.012}^{+0.022}$ & $0.93_{-0.04}^{+0.04}$ & \cite{2009MNRAS.394.2141N,2018MNRAS.478.1900S} \\
\hline
\multirow{6}*{UV/optical}&$\rm{J1152+0702}$ & $32.3_{-4.2}^{+4.8}$ & $0.998_{-0.032}^{+0.000}$ & \cite{2015MNRAS.446.3427C} \\
&$\rm{J1158-0322}$ & $31.6_{-3.4}^{+4.7}$ & $0.898_{-0.035}^{+0.036}$ & \cite{2015MNRAS.446.3427C} \\
&$\rm{J0941+0443}$ & $44.7_{-4.8}^{+5.4}$ & $0.998_{-0.032}^{+0.000}$ & \cite{2015MNRAS.446.3427C} \\
&$\rm{J0303+0027}$   & $63.1_{-6.9}^{+7.7}$ & $0.998_{-0.032}^{+0.000}$ & \cite{2015MNRAS.446.3427C} \\
&$\rm{J0927+0004}$& $15.8_{-2.0}^{+1.9}$ & $ 0.998_{-0.036}^{+0.000}$ & \cite{2015MNRAS.446.3427C} \\
\hline
\multirow{5}*{\shortstack{Most\\ massive}}&OJ 287 & $183.48\pm0.08$ & $0.381\pm0.004$ & \cite{2018ApJ...866...11D} \\
&Ton 618   & $660$ & $>0.60$ & \cite{2004ApJ...614..547S} \vspace{3mm} \\
&\shortstack{SDSS J140821.67\\ +025733.2}  & $1960$ & $>0.97$ & \cite{2017ApJS..228....9K} \\
\hline
\hline
\end{tabular}
\label{table:BH}
\end{table*}

\begin{figure}[htbp]
\subfigure[]{\includegraphics[width=1.05\columnwidth]{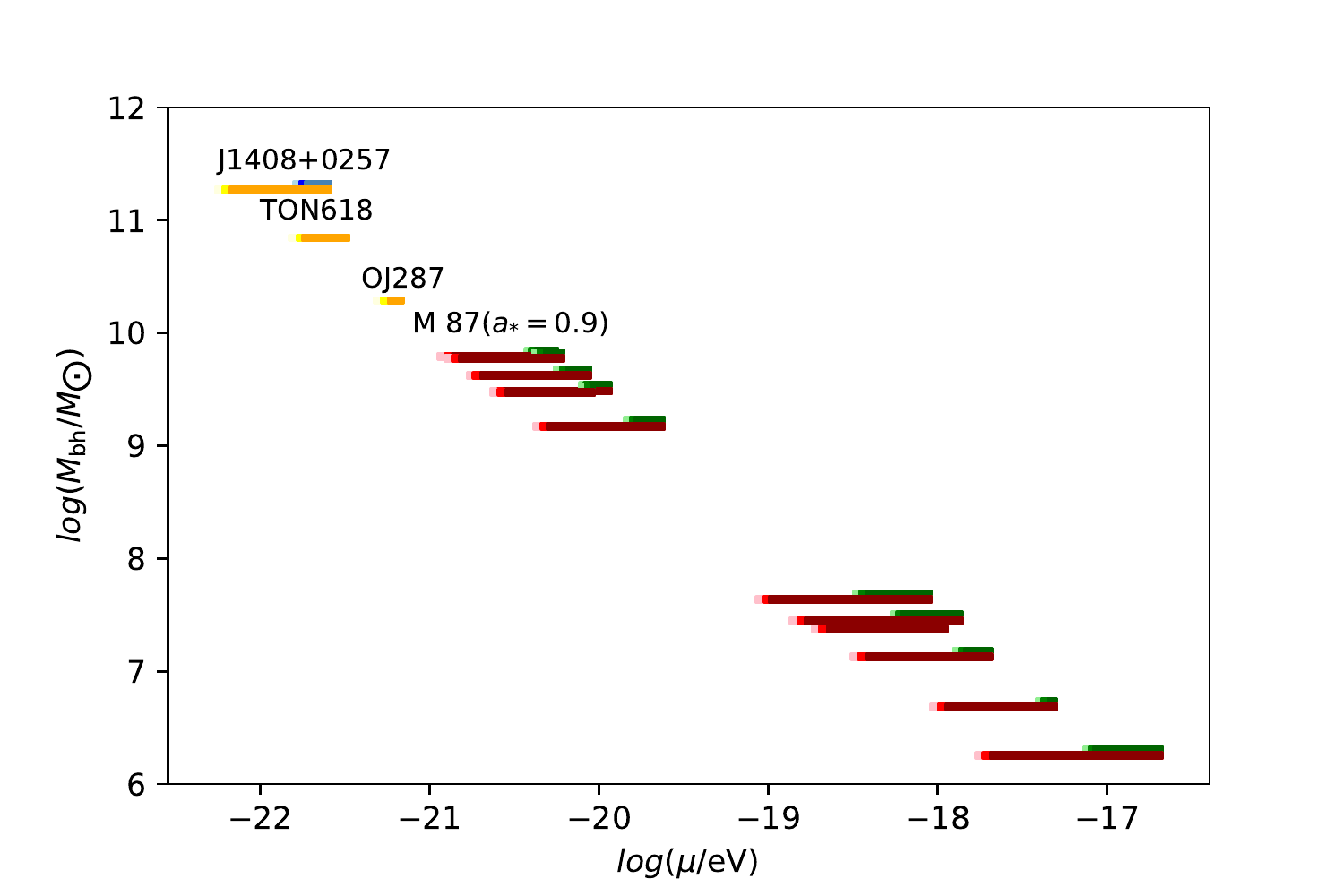}}
\subfigure[combined bounds]{\includegraphics[width=1\columnwidth]{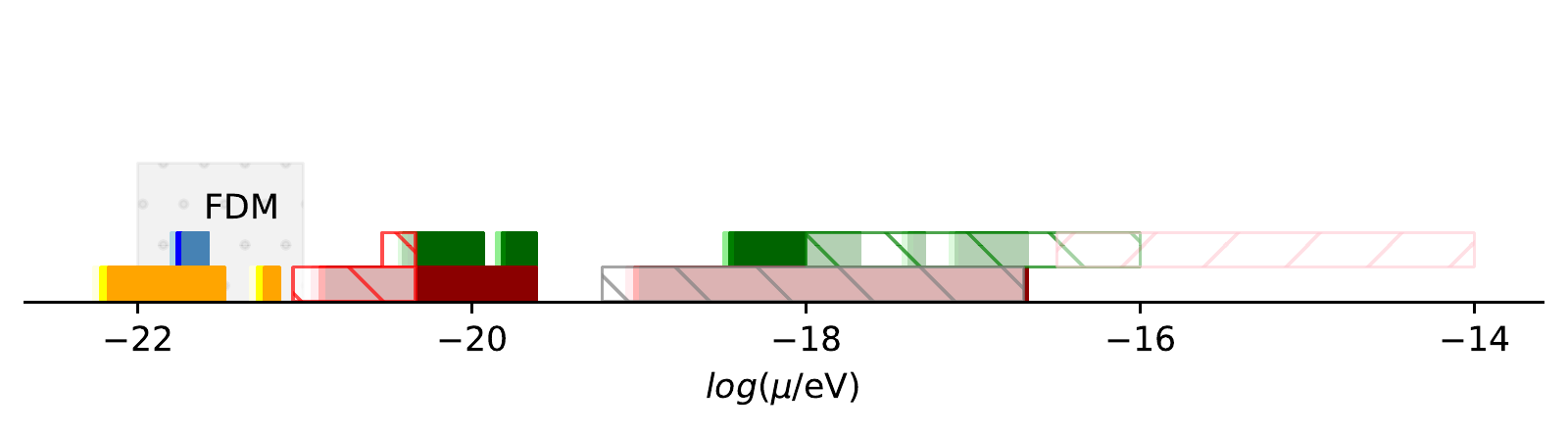}}
\caption{The light boson mass regions excluded by the SMBH samples
summarized in Table~\ref{table:BH}, together with those from M 87.
The green and red regions are for the cases of scalar and vector bosons,
respectively. For the most massive SMBHs, the corresponding bands are shown
in orange and blue. Here we assume $\tau_{\rm bh}=4.5 \times 10^7$ years.  The shaded green ($10^{-18} \rm{eV} \sim 10^{-16} $ eV), gray ($6 \times 10^{-20} \rm{eV} \sim 2 \times 10^{-11} $ eV) and red ($2.9 \times 10^{-21} \rm{eV} \sim 4.6 \times 10^{-21} \rm{eV}$ for scalars, and $8.5 \times 10^{-22} \rm{eV} \sim 4.6 \times 10^{-21} \rm{eV}$ for vectors) mass regions have been excluded in \cite{bokovi2019parametricresonance}, \cite{2017PhRvD..96c5019B} and \cite{2019PhRvL.123b1102D} by the SMBH superradiace effect. The pink mass regions ($10 ^{-16.5} \sim 10^{-14}$ eV) is expected to be probe by the LISA observations in the future \cite{Hannuksela:2018izj}.} 
\label{scalar_vector}
\end{figure}
\begin{figure*}[htbp]
\subfigure[SDSS J140821.67+025733.2]
{\includegraphics[width=1\columnwidth]{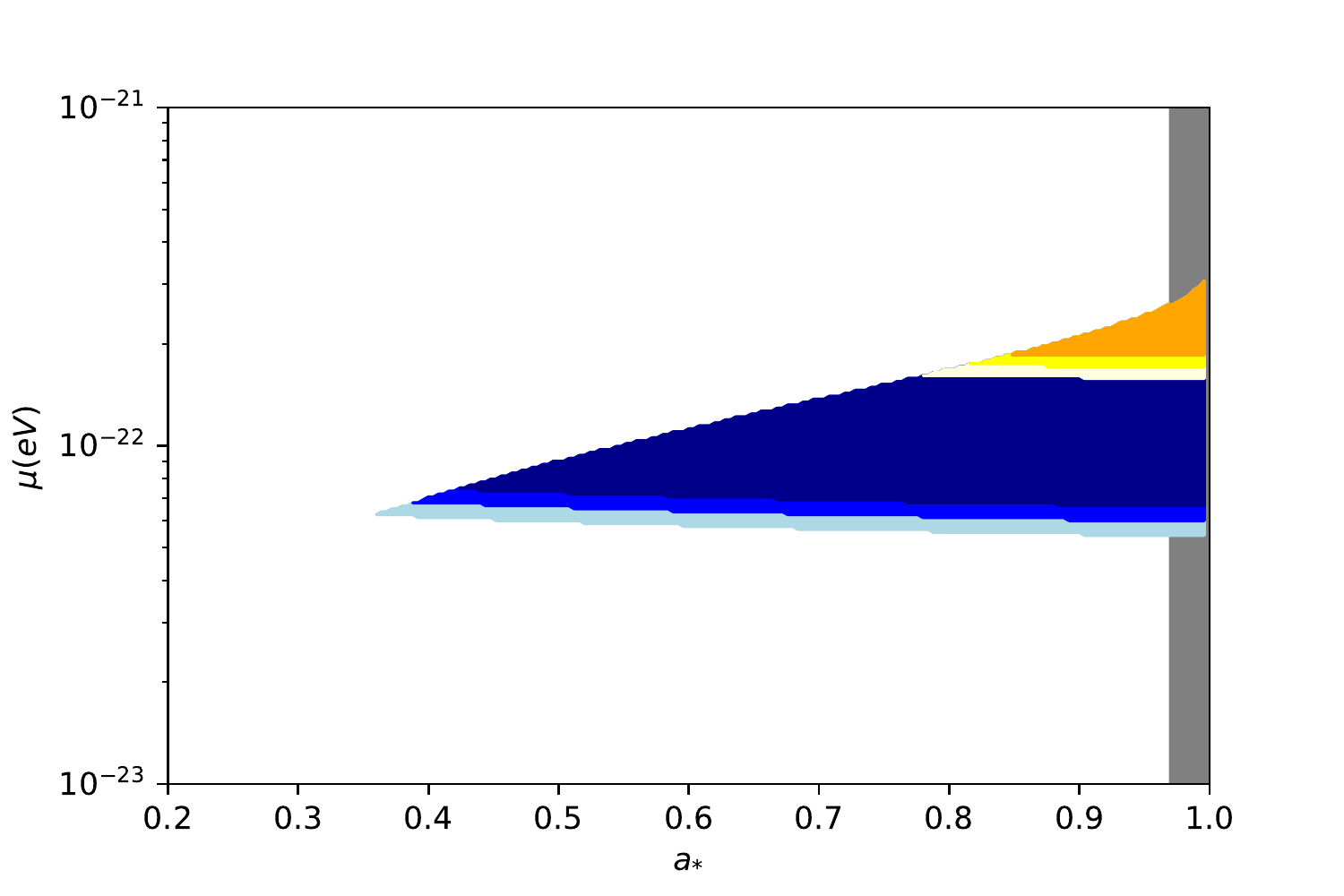}}
\subfigure[Ton 618]
{\includegraphics[width=1\columnwidth]{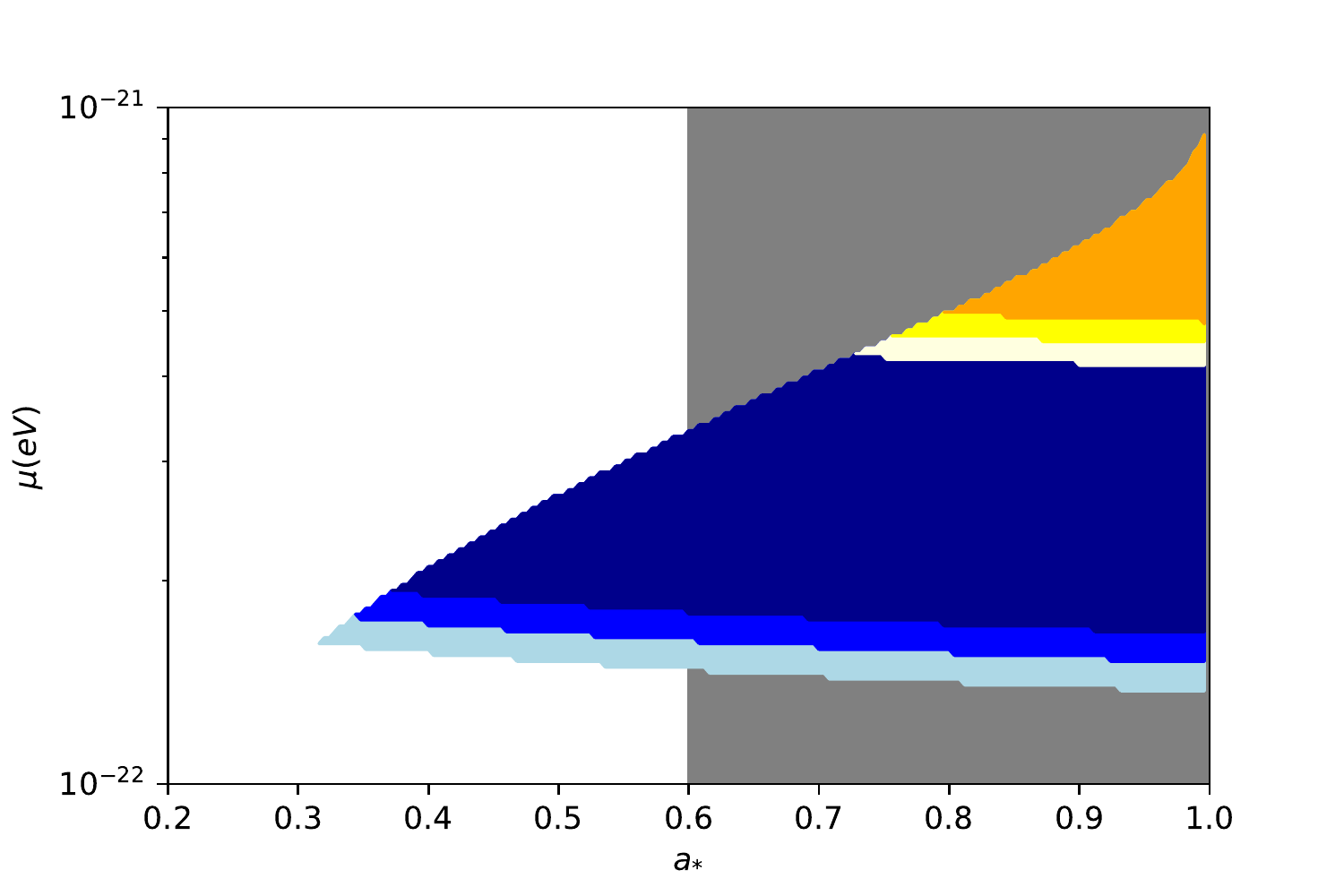}}
\caption{Constraints on the light boson mass as a function of the spin
of the BH, for SDSS J140821.67+025733.2 (a) and Ton 618 (b). The gray
regions show the allowed spins based on the accretion model of
Ref.~\cite{2016MNRAS.456L.109K}. The orange (blue) regions are those
for scalar (vector) bosons. Again we assume
$\tau_{\rm bh} = 4.5 \times 10^7 $ years.}
\label{mostmassivebh}
\end{figure*}

We firstly ignore the self-interaction of light bosons
\cite{2019PhRvL.123b1102D,2017PhRvD..96c5019B,2019arXiv191107862F}.
In this case the constraints can be
obtained through the combination of Eq.~(\ref{eq:1}) and Eq.~(\ref{eq:8}).
We further restrict our discussion in the $m=1$ mode of the boson cloud,
which is the dominant mode with the largest field strength. The exclusion
mass ranges for different sources are shown in Fig.~\ref{scalar_vector}.
The green (red) bands correspond to the exclusion regions for the scalar
(vector) bosons by the SMBHs summarized in Table~\ref{table:BH}.
The orange (blue) bands are those for the most massive SMBHs OJ 287,
Ton 618, and SDSS J140821.67+025733.2. Our results extend
remarkably the constraints compared with earlier approaches
\cite{2019PhRvL.123b1102D,2015PhRvD..91h4011A,2017PhRvD..96c5019B}.
Panel (b) of Fig.~\ref{scalar_vector}
summarizes the combined exclusion mass regions for all the SMBHs used in
this work. We can see that the mass range of $10^{-22} \sim 10^{-17}$~eV
can be effectively constrained by these sources, particularly for the vector
boson scenario due to its faster superradiance rate compared with the scalar
scenario (see Eq.~(\ref{eq:4}) and Eq.~(\ref{eq:5})). There is a lack of
constraints for $\mu\sim10^{-19}$~eV. We expect that additional SMBHs with
masses around $10^8$~M$_{\odot}$ will be helpful to close the mass window
around $10^{-19}$~eV. Fig.~\ref{mostmassivebh} shows the exclusion
capabilities of SDSS J140821.67+025733.2 and Ton 618 for a wide range of
spin. Even for $a_*\sim 0.4$, the constraints are still effective.

It is interesting to note that the most massive SMBHs OJ 287, SDSS
J140821.67+025733.2, and Ton 618 can exclude a good fraction of the region
of the so-called FDM model (with boson mass of $10^{-22}`` \sim 10^{-21}$~eV
\cite{2017PhRvD..95d3541H}) if it consists of vector bosons. For the scalar
case, the constraint is less stringent.

The exclusion regions also depend on the value of the BH lifetime
$\tau_{\rm bh}$. Longer lifetime would give wider exclusion regions.
In Fig.~\ref{scalar_vector}, we also show the constraints from M 87, with
$M_{\rm bh}=(6.5 \pm 0.7) \times 10^{9} M_{\odot}$ and $a_{*}=0.9 \pm 0.1$.
Here we take $\tau_{\rm bh}\sim 4.5\times 10^{7}$ years rather than
$\sim 10^{9}$ years as in Ref.~\cite{2019PhRvL.123b1102D}. This is why our
constraint is not as strong as that in Ref.~\cite{2019PhRvL.123b1102D} for
this specific object. Anyhow, our enlarged SMBH sample with diverse masses
and spins are clearly more powerful in constraining the light boson model.

 The differences between the results of our work and previous are also shown in Fig.~\ref{scalar_vector}. Our work have improved the constrains in the fuzzy dark matter mass range and the mass range about $\sim 10^{-20}$ eV.

\begin{figure}[htbp]
\centering
\includegraphics[width=1\columnwidth]{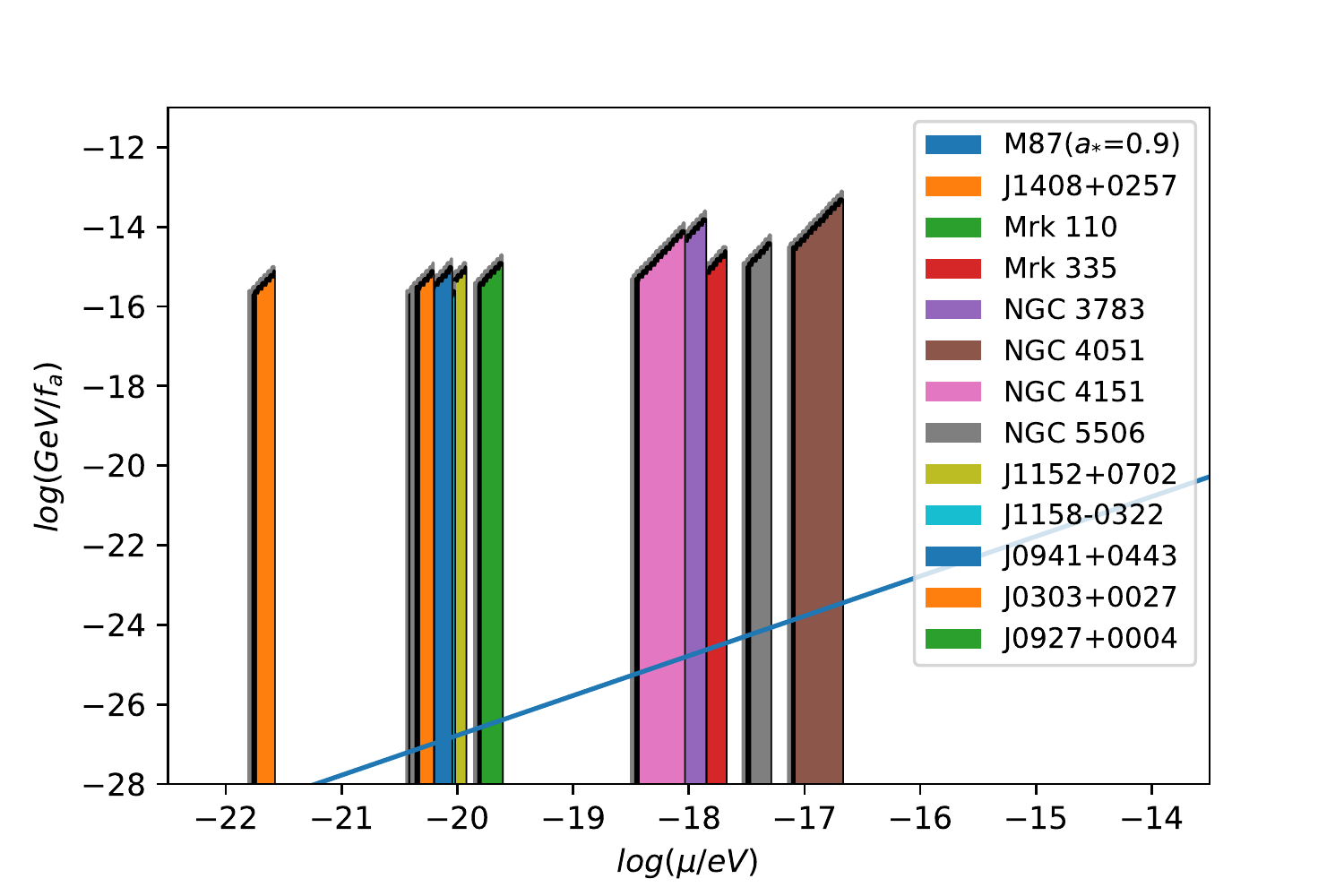}
\caption{Constraints on the mass and coupling of axions from quickly
rotating SMBHs (at 1$\sigma$ confidence level). The blue line is the
theoretical prediction for QCD axions \cite{2015PhRvD..91h4011A}.}
\label{superradiace}
\end{figure}

If the self-interaction is strong enough ($N_{\rm{bosenova}}<N_{\rm{max}}$),
the boson cloud collapses when growing up to the critical size
$N_{\rm{bosenova}}$, after which the superradiance process restarts and
the cycle repeats. The exclusion condition is now Eq.~(\ref{eq:1}) and
Eq.~(\ref{eq:10}). Generally speaking, including the self-interaction of
bosons would introduce one more free parameter, the decay constant $f_a$,
and relax somehow the constraints. As an illustration, we discuss the scalar
case in this work. The bosenova process for vector bosons with
self-interactions is more complicated, and may need more dedicated studies
in future. Here we use the numerical solution to the superradiance rate
$\Gamma$ for scalars such as axions in Ref.~\cite{2007PhRvD..76h4001D}.
The excluded parameter space in the mass-coupling plane of the scalars
is shown by the shaded regions in Fig.~\ref{superradiace}, assuming
$\tau_{\rm{bh}}=4.5 \times 10^7$~years. The mass range from $\sim 3 \times
10^{-19}$~eV to $10^{-17}$~eV with a decay constant $f_{a}>10^{15}$~GeV
can be excluded. It is very interesting to see that the theoretical
prediction of the mass-coupling relation for QCD axions
\cite{2015PhRvD..91h4011A}
has been excluded in a wide mass range with our sample. Note also that
the spin method constrains the axion parameters with large $f_a$, which
is complementary to the polarization method \cite{2020PhRvL.124f1102C}.

\section{Summary and discussion}

Superradiance leads to an effective extraction of angular momentum from
a rapidly-rotating BH. Therefore the SMBHs with measured masses and spins
serve as powerful probes of the presence of light bosons around rotating
BHs. Using a sample of high-spin SMBHs, inferred from the UV-optical or
X-ray spectroscopy, we constrain the model parameters of light bosons in
this work. The boson mass in the range of $(10^{-22} \sim 10^{-17})$~eV
are effectively constrained. The most massive BHs OJ 287, SDSS
J140821.67+025733.2, and Ton 618 can effectively extend the constraints
to the FDM mass range. We also consider to include the self-interactions
for scalar bosons, and exclude a large part of the parameter space for
$3 \times 10^{-19}<\mu/{\rm eV}<10^{-17}$ and $f_{a} > 10^{15}$~GeV.

We are aware that there are several uncertainties of the results, such
as the systematical uncertainties of the parameter estimates of the BH
masses and spins, and the calculation of the superradiance rate of bosons.
Furthermore, the inclusion of other modes of the boson cloud would also
change somehow the quantitative results derived in this work. We leave
such improvements/refinements in future works. We expect that more
precise determinations of the masses and spins of SMBHs in future will
provide significantly improved and robust constraints on the light bosons
in a wide mass region.



\section*{Acknowledgements}
This work is supported by the National Key Research
and Development Program of China (Grant No. 2016YFA0400200), the National
Natural Science Foundation of China (Grants No. 11525313, No. 11722328,
No. 11773075, No. U1738210, No. U1738136, and No. U1738206), the 100
Talents Program of Chinese Academy of Sciences, and the Youth Innovation
Promotion Association of Chinese Academy of Sciences (Grant No. 2016288).

\bibliographystyle{apsrev}
\bibliography{refs}

\end{document}